\documentclass{article}
\usepackage{spconf,amsmath,graphicx}
\usepackage{amsmath}
\usepackage{amsthm}
\usepackage{algorithm}
\usepackage{algorithmic}
\usepackage{colortbl}
\usepackage{multirow}
\usepackage[switch]{lineno}

\usepackage{xspace}
\usepackage{booktabs}  
\usepackage{threeparttable}  
\usepackage{multicol}  
\usepackage{multirow}  
\usepackage{mathrsfs}
\usepackage{stfloats}
\usepackage{times}  
\usepackage{helvet}  
\usepackage{courier}  
\usepackage[hyphens]{url}  
\usepackage{graphicx} 
\usepackage{amsmath}
\usepackage{amssymb}
\usepackage{stmaryrd}
\usepackage{arydshln}

\usepackage[switch]{lineno}
\usepackage{xspace}
\usepackage{colortbl} 
\usepackage{multicol}  
\usepackage{multirow}  
\usepackage{mathrsfs}
\usepackage{stfloats}

\usepackage{times}
\usepackage{soul}
\usepackage{url}
\usepackage[hidelinks]{hyperref}
\usepackage[utf8]{inputenc}

\usepackage{graphicx}
\usepackage[dvipsnames]{xcolor}
\usepackage{amsmath}
\usepackage{amsthm}
\usepackage{amsfonts}
\usepackage{booktabs}
\usepackage{graphicx}
\usepackage{bm}
\usepackage{bigstrut}
\renewcommand{\paragraph}[1]{\noindent\textbf{#1}\quad}

\title{TransAudio: Towards the Transferable Adversarial Audio Attack via Learning Contextualized Perturbations}
\name{%
\begin{tabular}{@{}c@{}}
Gege Qi \textsuperscript{\rm 1} \qquad 
Yuefeng Chen \textsuperscript{\rm 1} \qquad 
Xiaofeng Mao \textsuperscript{\rm 1} \qquad
Yao Zhu \textsuperscript{\rm 2} \\
\textit{Binyuan Hui\textsuperscript{\rm 1}} \qquad 
\textit{Xiaodan Li\textsuperscript{\rm 1}} \qquad 
\textit{Rong Zhang\textsuperscript{\rm 1}} \qquad 
\textit{Hui Xue\textsuperscript{\rm 1}} 
\end{tabular}}

\address{\textsuperscript{\rm 1}Alibaba Group, Hangzhou, China \textsuperscript{\rm 2}Zhejiang University, Hangzhou, China \\
 \textrm{\{qigege.qgg,yuefeng.chenyf,mxf164419\}@alibaba-inc.com}}

\begin{document}
%
\maketitle
\begin{abstract}
In a transfer-based attack against Automatic Speech Recognition (ASR) systems, attacks are unable to access the architecture and parameters of the target model.
Existing attack methods are mostly investigated in voice assistant scenarios with restricted voice commands, prohibiting their applicability to more general ASR related applications.
To tackle this challenge, we propose a novel contextualized attack with deletion, insertion, and substitution adversarial behaviors, namely TransAudio, which achieves arbitrary word-level attacks based on the proposed two-stage framework. To strengthen the attack transferability, we further introduce an audio score-matching optimization strategy to regularize the training process, which mitigates adversarial example over-fitting to the surrogate model.
Extensive experiments and analysis demonstrate the effectiveness of TransAudio against open-source ASR models and commercial APIs. 
\end{abstract}
\section{Introduction}

\label{sec:intro}
Automatic Speech Recognition (ASR) systems have been integrated into numerous intelligent voice control agents, like Amazon Alex, opening up new ways for humans to interact with ubiquitous smart devices. However, the free and comfortable interaction poses potential privacy and security risks if attacked with malicious intent. Recent works have shown that audio Adversarial Examples (AEs) can lead ASR systems to malfunction, which might cause unfavorable outcomes \cite{carlini2018audio,zhang2021commandergabble,wang2020adversarial,khare2018adversarial}.
\let\thefootnote\relax\footnotetext{This research is supported in part by the National Key Research and Development Program of China under Grant No.2020AAA0140000.}

\begin{figure}
    \centering
    \includegraphics[width=0.48\textwidth]{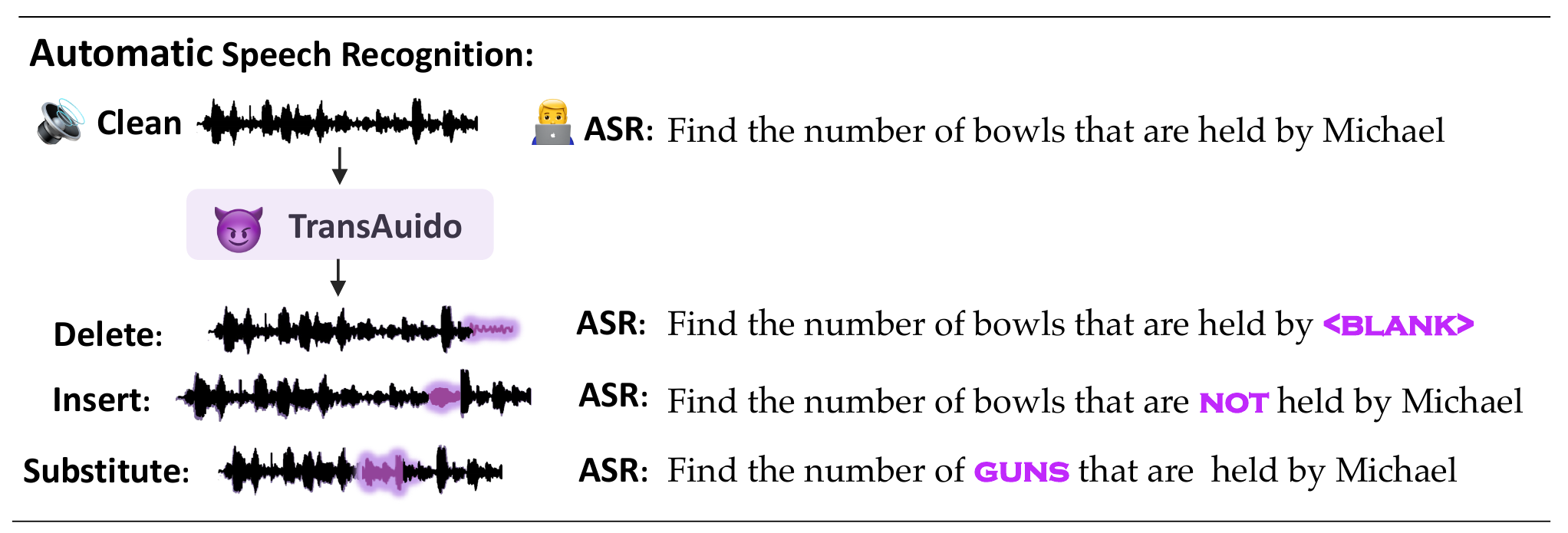}
    \vspace{-0.5cm}
    \caption{Transfer attack with three contextualized perturbations: Delete, Insert and Substitute.}
    \vspace{-0.7cm}
    \label{fig:ito}
\end{figure}


Most previous adversarial attacks \cite{carlini2018audio, yuan2018commandersong, esmaeilpour2022towards} consider the white-box setting, where the target model is fully exposed to the attacker. 
In contrast to having a full model, which is impractical for an attacker, a black-box attack is more attractive since only the input and output of the target model are used to accomplish the attack.
Modern black-box attacks can be divided into query-based and transfer-based approaches. The query-based attacks \cite{zheng2021black} estimate the direction of the gradient through a large number of adversarial example inferences, which are effective but costly as the number of accesses (known as query budgets) to the target model may be limited.
Fortunately, the transfer-based approach \cite{chen2020devil} shines a light on how to use the surrogate model to generate transferable AEs to attack target models, eliminating the cost of query budget. 
However, we observed that the current transfer-based ASR attack still has two major limitations, including \textbf{the limited attack space of target commands} and \textbf{easily over-fitting to the surrogate model}.

\textbf{Firstly}, existing ASR attacks only consider a limited set of short commands, e.g., \texttt{[turn light on]} and \texttt{[clear notification]}. They are effective in a narrow attack space with a complexity of $\mathcal{O}(C)$, where $C$ is the number of \textbf{C}ommands, which prevents application to general real-time ASR systems.
Motivated by text attack \cite{li2020contextualized}, we consider that a realistic ASR attack can alter the meaning of audio via word-level perturbation in context, as shown in Fig.~\ref{fig:ito}, including substitution, insertion, and deletion contextualized attacks, rather than commands-level attacks.
This makes the attacker to extend the attack space from $\mathcal{O}(C)$ to $\mathcal{O}(VL)$, where $V$ is size of the \textbf{V}ocabulary and $L$ is the text \textbf{L}ength and $VL \gg C$. 
However, we observed that existing methods are extremely hard to perform arbitrary word-level targeting attacks when trivially trying to grow the size of the target space.
Presumably, the ASR needs to merge the sequential character predictions and produce a sentence where the distorted temporal dependency of AEs can not be controlled well for the purpose of targeted attacks. 

\textbf{Secondly}, over-fitting of surrogate models is commonly seen as an obstacle to adversarial transfer attacks \cite{xie2019improving}, yet there is a lack of an effective regularization strategy to mitigate this problem.

In this paper, we propose a two-stage attack framework \textsc{TransAudio} that can perform arbitrary word-level attacks with high transferability.
Concretely, the first stage performs a local attack, where the AE fragment for the target word is generated via text-to-speech (TTS) models.
The second stage performs a global attack that optimizes the contextualized global perturbations on entire sequences composed of a local AE fragment and unconcerned areas of the input audio.
Besides, we further propose a score-matching based optimization strategy for the surrogate model. 
By introducing data distribution information into surrogate model, such a distribution alignment method can lessen the dependence of the perturbation on the surrogate model and enhance the adversarial transferability.



We show the effectiveness of TransAudio on both Chinese and English speech recognition systems.
Extensive experiments on open-source ASR models show that the contextualized adversarial examples outperform traditional attacks in terms of attack success rate, character error rate, and edit distance. 
Notably, TransAudio can successfully generate adversarial examples on large target spaces when launching targeted attacks toward commercial APIs.

\begin{figure}
    \centering
    \includegraphics[width=0.48\textwidth]{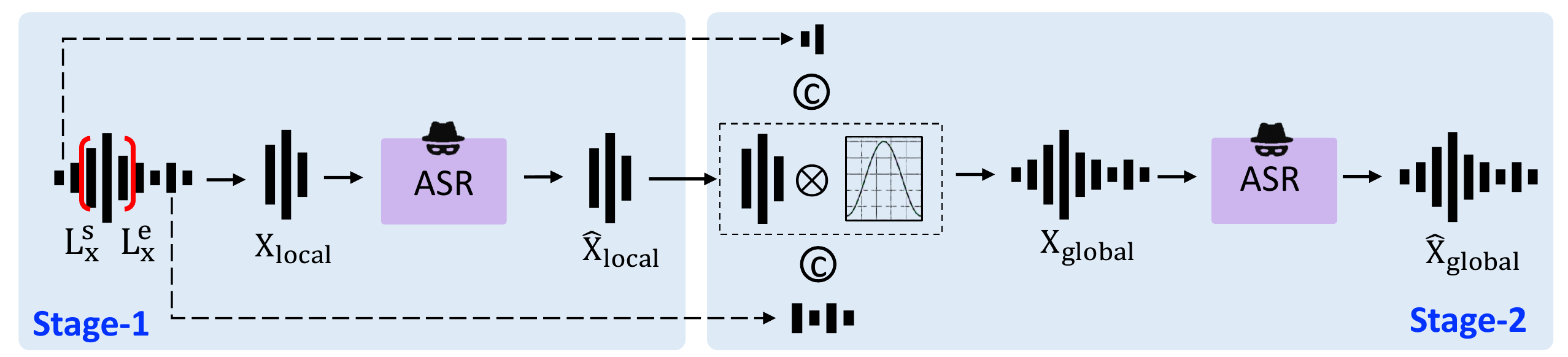}
    \vspace{-0.5cm}
    \caption{Overview of TransAudio. Stage-1 crafts a local AE of specific target word, which is fed into Stage-2 for contextualized global audio AE generating.}
    \vspace{-0.7cm}
    \label{fig:TransAudio}
\end{figure}

\section{Problem Formulation}
Given a clean audio $X$ with text label $Y$, the object of an end-to-end ASR is to model the joint probability of a $M$-length output sequence $Y=\left( y_1, ..., y_M | y_i \in \mathcal{V} \right)$ given a $N$-length input sequence $X=\left(x_1,...,x_N | x_i \in \mathbb{R}^D \right)$. Here, $y_i$ is the $i$-th word in the output sequence $Y$.
Then, the targeted adversarial attacks associated with deletion, insertion, and substitution can be described as the procedure of optimizing an AE such that it deceives the ASR to the target text. Denote $\hat{Y}_k$ as the target label formed by modifying the $k$-th word in $Y$. Then $\hat{Y}_k$ can be expressed as:
\begin{equation}
\begin{small}
    \hat{Y}_k=
    \begin{cases}
      \left(y_1,..., y_{k-1}, y_{k+1}, ..., y_M\right)  & \text{Delete} \\
      \left(y_1,..., y_k, \hat{y}_k, y_{k+1}, ..., y_M\right) & \text{Insert} \\
      \left(y_1,..., y_{k-1}, \hat{y}_k, y_{k+1}, ..., y_M\right) & \text{Substitute} 
    \end{cases}
\end{small}
\end{equation}
where $\hat{y}_k$ is the target word. With the above target labels, the task of finding audio AEs can be rephrased as solving a constrained continuous problem:
\begin{equation}
\begin{small}
 \min \mathcal{J}(\hat{X}, \hat{Y}_k), \mathtt{s.t.} ||\hat{X} - X||_p \leq \delta,
\end{small}
\end{equation}
To attack the ASR model which is developed based on CTC-attention encoder-decoder architectures \cite{kim2017joint}, accordingly, the object function $\mathcal{J}$ is as follows:
\begin{equation}\label{eq:loss}
\begin{small}
\mathcal{J_{ASR}} = \lambda \mathcal{J}_{ctc}(\hat{X},\hat{Y}_k) + (1-\lambda) \mathcal{J}_{att}(\hat{X},\hat{Y}_k).
\end{small}
\end{equation}

where $\mathcal{J}_{ctc}$ is the connectionist temporal classification loss, $\mathcal{J}_{att}$ is the attention loss and $\lambda$ is a tunable parameter.

\section{Proposed Approach: TransAudio}
\label{sec:method}



\subsection{Two-stage Attack}
To craft canonical and effective AEs, we adopt a two-stage strategy to conduct the contextualized adversary, as shown in Fig.~\ref{fig:TransAudio}, which can be seen as a local-to-global process.




\noindent\textbf{First Stage: local attack.} 
TransAdudio first applies a local audio attack corresponding to the target word.
The prerequisite for an arbitrary word-level attack is to locate the target place of the attack in the audio. 
We propose an audio location module (ALM) to determine the target location, implemented by CT-Segmentation~\cite{watanabe2018espnet}. 
Formally, given an audio $X$ with target word $y_k$ to be attacked, the start and end position of the original audio signal could be produced by:
\begin{equation}\label{eq:ALM}
\begin{small}
[L_x^{s},L_x^{e}] = ALM(X,y_k).
\end{small}
\end{equation}
Then, the local adversarial audio $\hat{\mathbf{x}}_{local}$ can be generated by the iterative attack method \cite{dong2018boosting} on surrogate models, according to location $[L_x^{s},L_x^{e}]$.
To generate transferability-effective audio AEs, for insertion and substitution attacks, we first initialize $\hat{\mathbf{x}}_{local}$ with a natural audio synthesized by Text-To-Speech (TTS) models.
In practice, we turn to the end-to-end TTS model \cite{ren2020fastspeech} that could easily synthesize waveform given text phrases $\hat{y}_k$ as semantic contents.
Furthermore, based on stealthy considerations, we apply a high-pass filter $F_h(\cdot)$ at a cutoff frequency of 7kHz resulting $\hat{\mathbf{x}}_k = F_h(TTS(\hat{y}_k))$ and thus less detectable by the human ear. 
Formally, the initialized local AE $\hat{\mathbf{x}}_{local}^0$ can be represented as follows:
\begin{equation}
\begin{small}
    \label{eq:x_local_init}
    \hat{\mathbf{x}}_{local}^0=
    \begin{cases}
      X[L_x^s : L_x^e]  & \text{Delete} \\
      \hat{\mathbf{x}}_k & \text{Insert} \\
       \frac{\hat{\mathbf{x}}_k + X[L_x^s : L_x^e]}{2} & \text{Substitute} 
    \end{cases}
\end{small}
\end{equation}
where the local adversary $\hat{\mathbf{x}}_{local}^0$ is initialized as original audio fragment $X[L_x^s : L_x^e]$ for deletion attacks. As for insertion attacks, $\hat{\mathbf{x}}_{local}^0$ has to be clipped by $\mathtt{clip}$ operation with the maximum value of $X$. To hide the target information in the original audio, in the substitution scenario, we enrich the local audio fragment $X[L_x^s : L_x^e]$ by adding the synthesized target audio $\hat{\mathbf{x}}_k$, and then clip it with budget $\delta$.
And after that, the MI-FGSM \cite{dong2018boosting} method is used for generating local audio AEs based on the gradient information of source models.
Formally, the equation for generating audio AEs is as follows:
\begin{equation}
\begin{small}
    \begin{cases}
      g_{n+1} &= \mu \cdot g_{n} + \frac{\nabla_{\mathbf{x}}\mathcal{J}(\hat{\mathbf{x}}^{n}_{local},\hat{y}_{k}))}{|| \nabla_\mathbf{x}\mathcal{J}(\hat{\mathbf{x}}^{n}_{local},\hat{y}_{k})) ||_{1}},  \\
\hat{\mathbf{x}}^{n+1}_{local} &= \mathtt{clip}(\hat{\mathbf{x}}^{n}_{local} - \alpha \cdot \mathtt{sign}(g_{n+1}))
    \end{cases}
\label{eq:mi-fgsm-yf}
\end{small}
\end{equation}
where $\mu$ is the decay factor of the momentum term, $\alpha$ is the step size, and $g_n$ is the accumulated gradient at iteration $n$. 

\noindent\textbf{Second Stage: global attack.} 
Then, we propose to conduct a global attack to obtain final AEs with the global target $\hat{Y}_k$.
The initialized global AE can be formed by concatenating the local AE and the remaining audio sequence. However, we find that this simple stitching operation causes a marked deformation of transcription by the target ASR. 
To address this, we add a hamming window $w(\cdot)$ to each waveform fragment for non-interference of signals.
Thus, the initialized global AE can be formed as $\hat{X}^{0}_{global} = \mathtt{Concat}(w(X[:L_x^{s}]), w(\hat{\mathbf{x}}_{local}), w(X[L_x^{e}:]))$.

Second stage also solves the optimization problem in a gradient descent manner, which only needs to replace $\hat{y}_k \rightarrow \hat{Y}_k, \hat{\mathbf{x}}^{n}_{local} \rightarrow \hat{X}^{n}_{global} $ in Equation~\ref{eq:mi-fgsm-yf}. Noted that, in the insertion attack, $\hat{X}^{0}_{global} = \mathtt{Concat}(w(X[:L_x^{s}]), w(\hat{\mathbf{x}}_{local}), w(X[L_x^{s}:]))$, i.e., inserting local AEs at specific locations $L_x^{s}$, while the other two scenarios use the replacement operation.

\begin{algorithm}[t]
\caption{TransAudio Attack}
\label{alg:TransAudio}
\textbf{Input}: ASR $f_{\theta}$, loss function $\mathcal{J}$, hamming function $w(\cdot)$, the original audio $X$, local targeted audio $\hat{\mathbf{x}}_{local}$, local targeted word $\hat{y}_k$, global targeted sentence $\hat{Y}_k$, location of the targeted word $[L_x^{s},L_x^{e}]$, the size of perturbation $\delta$, total iterations $T$, decay factor $\mu$, attack type $A_{type}\in \left\{ins, del, sub\right\}$. \\
\textbf{Output}: The adversarial audio sample $\hat{X}_{global}$. \\
\textbf{First-stage}: Generate local audio AEs. \\
\vspace{-4mm}\begin{algorithmic}[1]
\STATE Let $i=0; g_{0}=0; \hat{\mathbf{x}}_{local}^0=\mathtt{clip}(\hat{\mathbf{x}}_k +X[L_x^s : L_x^e])$
\WHILE{$i<\frac{T}{2}$}
\STATE \textit{\{Calculate local audio AEs $\hat{\mathbf{x}}_{local}^i$\}}
$\triangleright$ {Equation~\ref{eq:mi-fgsm-yf}} 
\ENDWHILE
\STATE \textbf{return} $\hat{\mathbf{x}}_{local} = \hat{\mathbf{x}}_{local}^{\frac{T}{2}}$
\end{algorithmic}

\textbf{Second-stage:} {Generate global audio AEs.} \\
\vspace{-4mm}\begin{algorithmic}[1]
\STATE Let $i=0; g_{0}=0; $
\STATE $\hat{X}^{0}_{global} = \mathtt{Concat}(w(X[:L_x^{s}]), w(\hat{\mathbf{x}}_{local}), w(X[L_x^{e}:]))$

\WHILE{$i<\frac{T}{2}$}
\STATE \textit{\{Calculate global audio AEs $\hat{X}^{i}_{global}$ by replacing $\hat{y}_k \rightarrow \hat{Y}_k, \hat{\mathbf{x}}^{n}_{local} \rightarrow \hat{X}^{n}_{global}$\}} $\triangleright$ {Equation~\ref{eq:mi-fgsm-yf}}
\ENDWHILE
\STATE \textbf{return} $\hat{X}_{global}=\hat{X}^{\frac{T}{2}}_{global}$
\end{algorithmic}
\end{algorithm}

\subsection{Audio Score-Matching based Surrogate Model}
We focus on utilizing the gradient of a fixed surrogate model to obtain a transferable AE. Inspired by the image transferable attack \cite{zhu2022towards}, we can know that pushing the input away from its original distribution can enhance the adversarial transferability. Thus, we propose a score-matching based optimization strategy for audio attack, aiming to enforce the direction of the model’s input-gradients update towards gradients of the data distribution. 
We utilize the sequence of discrete output tokens of the surrogate ASRs, i.e., the ground truth of $t$-th token $z_t$, to minimize the distance between the gradient of log conditional density $p_{\theta}{(z_t|X)}$ and the gradient of log ground truth class-conditional data distribution $p_{D}{(X|z_t)}$ as follows:
\begin{gather}
\begin{footnotesize}
\label{eq:sm1}
   \begin{split}
    \mathcal{J}_{ASM}^{t} & = \mathbb{E}_{p_{D}(z_t)}\mathbb{E}_{p_{D}(x|z_t)}||\nabla_{X}\log p_{\theta}(z_t|X)-\nabla_{X}\log p_{D}(X|z_t)||_2^2 \\
    & \approx \mathbb{E}_{p_{D}(z_t)}\mathbb{E}_{p_{D}(x|z_t)}||\nabla_{X}\log p_{\theta}(z_t|X)||_2^2 \\
    & +2\cdot\mathbb{E}_{p_{D}(z_t)}\mathbb{E}_{p_{D}(x|z_t)}[tr(\nabla^{2}\log p_{\theta}(z_t|X))]
\end{split}
\end{footnotesize}
\end{gather}
where $\operatorname{tr}(\cdot)$ is the matrix trace.
Due to the high audio sampling rate (e.g., 16kHz), it is hard to calculate the trace of Hessian matrix in the second term of Equation~\ref{eq:sm1}. Following \cite{pang2020efficient}, we apply the Hutchinson's trick to approximate $tr(A)$ as: $tr(A)=\mathbb{E}_{p(\nu)}[\nu^{T}A\nu]$, where $\nu$ is a random vector thus that $\mathbb{E}_{p(\nu)}[\nu^{T}\nu]=I$. To this end, the Audio Score-Matching (ASM) loss can be written as:
\begin{gather}\label{eq:fd-dsm}
\begin{small}
\begin{split}
\mathcal{J}_{ASM} & = \frac{1}{M}\sum_{t=1}^M [\mathbb{E}_{p_{D}(z_t)}\mathbb{E}_{p_{D} (x|z_t)}||\nabla_{X}\log p_{\theta}(z_t|X)||_2^2 \\
& + 2 \cdot \mathbb{E}_{p_{D}(z_t)}\mathbb{E}_{p_{D}(x|z_t)}\mathbb{E}_{p(\nu)}[\nu^{T}\nabla_{X}^2\log p_{\theta}(z_t|X)\nu ]]
\end{split}
\end{small}
\end{gather}

We fine-tune the surrogate model by optimizing the ASR loss described in Equation~\ref{eq:loss} and ASM loss jointly. This yields the following objective:
\begin{equation}\label{eq:fd-dsm3}
\begin{small}
\mathcal{J}_{Total} = \mathcal{J}_{ASR} + \lambda_{ASM}\mathcal{J}_{ASM}
\end{small}
\end{equation}%
where we set $\lambda_{ASM}=0.01$. With the fine-tuned surrogate model, the transferability of AEs can be enhanced by modifying the input through its integrated gradients, i.e., towards the gradient of real data distribution. Note that TransAudio$^*$ is denoted as the TransAudio with the ASM-surrogate model. 

\begin{table*}
\centering
\small
\renewcommand{\arraystretch}{1.15}
\resizebox{0.8\hsize}{!}{
\begin{tabular}{l | c | cccc|cccc|cccc}
\hline
 \multirow{2}*{\textbf{\textsc{Dataset}}}& 
\multirow{2}*{\textbf{\textsc{Method}}} & \multicolumn{4}{c|}{\textbf{\textsc{Deletion}}}  & \multicolumn{4}{c|}{\textbf{\textsc{Insertion}}}   & \multicolumn{4}{c}{\textbf{\textsc{Subsitution}}}   \\
    ~ & ~ & SNR$\uparrow$ & SRoA$\uparrow$& CER$\downarrow$ & MED$\downarrow$ & SNR$\uparrow$ &SRoA$\uparrow$& CER$\downarrow$ & MED$\downarrow$ & SNR$\uparrow$ & SRoA$\uparrow$& CER$\downarrow$ & MED$\downarrow$ \\
\hline
 \multirow{4}{*}{AISHEEL} &Devil’s Whisper~\cite{chen2020devil} & 8.23 & 49 & 0.37 & 7.26 & 1.45 & 9 & 0.26 & 5.28 & -2.28 & 4 & 0.45 & 9.63 \\
~ &NI-OCCAM~\cite{zheng2021black} & 8.07 & 76 & 0.34 & 5.11 & 1.43 & 13 & 0.32 & 5.18 & -2.13 & 8 & 0.46 & 7.85  \\
~&TransAudio &15.02 & 87 & 0.30 & 4.37 & 1.43 & 26 & 0.31 & 5.22 & \textbf{-0.88} & 23 & 0.37 & 5.55 \\
~&TransAudio* &\textbf{15.61} & \textbf{91} & \textbf{0.24} & \textbf{3.29} 
 	 & \textbf{2.21} & \textbf{27} & \textbf{0.21} & \textbf{3.25} 
 	 & -0.9 & \textbf{26} & \textbf{0.35} & \textbf{5.42} \\
\hline
\hline
\multirow{4}{*}{Librispeech} & Devil’s Whisper~\cite{chen2020devil} & 5.38 & 27 & 0.31 & 11.52 & 1.37 & 16 & 0.22 & 9.71 & 4.56 & 14 & 0.59 & 10.04   \\ 
~ & NI-OCCAM~\cite{zheng2021black} & 5.06 & 44 & 0.14 & 10.23 & 1.41 & 36 & 0.10 & 7.54 & 6.63 & 27 & 0.13 & 8.86 \\
~ &TransAudio &5.45 & 61 & 0.09 & \textbf{6.88}
 	 & 1.41 & 48 & 0.06 & 4.96 
 	 & 7.51 & 27 & 0.29 & \textbf{6.71}  \\
~ & TransAudio* &\textbf{5.45} & \textbf{62} & \textbf{0.09} & 7.09 
 	 & \textbf{1.79} & \textbf{51} & \textbf{0.05} & \textbf{4.66}
 	 & \textbf{7.56} & \textbf{31} & \textbf{0.19} & 10.18\\
\hline
\end{tabular}}
\vspace{-0.2cm}
\caption{The results of contextualized attacks for English and Chinese audios on open-source ASRs.}
\label{tab:tab1}
\vspace{-0.5cm}
\end{table*}

\section{Experiments}
\paragraph{Datasets and Settings.}
We conduct experiments on AISHELL~\cite{bu2017aishell} and Librispeech \cite{panayotov2015librispeech} datasets.
AISHELL is an open-source Mandarin speech corpus re-sampled to 16 kHz. In deletion attacks, we use the Chinese text segmentation module\footnote{\url{https://github.com/fxsjy/jieba}} to segment the texts into words and randomly select the target words. As for insertion attacks, we randomly select the targeted words from the candidate of frequently used Chinese words. Librispeech corpus is derived from English audio books containing 1000 hours of speech. Since English words are separated by $<\mathtt{blank}>$, it is convenient to select the target words for deletion attack.
The format of audios is 16-bit WAV with 16 kHz, which is the mainstream setup for commercial products. 
We randomly select 100 audios in the test sets of AISHELL and Librispeech for evaluation. The perturbation budget $\delta$ is set as 0.06 under ${\ell_\infty}$-norm, the number of iteration $T$ in Algorithm~\ref{alg:TransAudio} is set to 50 and decay factor $\mu$ is 0.5.

\paragraph{Evaluation Metrics.}~We use the success rate of attack (SRoA) to evaluate the effectiveness of AEs.
Specific to deletion attacks, SRoA means that the target word can escape from ASRs. On the contrary, while insertion and substitution attacks aim to ensure recognition of the target words.
Following \cite{chen2020devil}, Character Error Rate (CER) and Minimum Edit Distance (MED) are two metrics we used to measure the impact of contextualized perturbations on the unconcerned audio regions. We use Signal-to-Noise Ratio (SNR) to evaluate the quality of adversarial audios.
For insertion attacks, we insert zero phonemes at the target location to align the AE's length with the original audio.


\paragraph{Target and Surrogate Models.}~We use Algorithm~\ref{alg:TransAudio} to attack open-source ASR models implemented on ESPNet \cite{watanabe2018espnet} and popular speech transcription APIs. Specifically, the target and surrogate models for AISHELL are conformer and streaming conformer. The target and surrogate models on Librispeech are conformer6 and transformer2 respectively. To demonstrate our attack in a truly black-box scenario, we attack the speech transcription APIs, including Google Cloud Speech-to-Text\footnote{\url{https://cloud.google.com/speech-to-text}}, Alibaba Short Speech Recognition \footnote{\url{https://www.alibabacloud.com/zh/product/intelligent-speech-interaction}} and Tencent Short Speech Recognition \footnote{\url{https://cloud.tencent.com/product/asr}}.


\paragraph{Evaluation on Open-Source ASR systems.}~Table \ref{tab:tab1} displays the performance of TransAudio and other competitive transfer targeted attacks on AISHELL and Librispeech. First, we demonstrate that TransAudio* achieves state-of-the-art on this challenging word-level attack. Also, it is evident that the SRoA of our method is vastly superior to other methods on insertion and substitution attacks, e.g., 15\% and 4\% higher than the NI-OCCAM in the English scenario.
The lower CER and MED of TransAudio* demonstrate that the proposed two-stage attack strategy has a reduced impact on the transcription of unconcerned fragments, which improves the stealthy of word-level attacks.
Moreover, we see a progressive decline in the SRoA of deletion, insertion, and substitution attacks. 
This is reasonable because the insertion attack introduces more additional information than the deletion attack and is strictly harder. Furthermore, substitution attempts to replace the original information, making an attack more challenging.
Notably, when compared with TransAudio, TransAudio$^*$ can further increase SRoA. This indicates that distribution-related information of AEs, which is guided by target text, helps to improve attack transferability in the black-box setting. 

\begin{table}[t]  
    \small
    \centering
    \resizebox{0.85\hsize}{!}{
    \begin{tabular}{cccccccc}  
    \toprule
    \multirow{2}{*}{\textbf{\textsc{APIs}}} & \multirow{2}{*}{\textbf{\textsc{Method}}}  &
    \multicolumn{2}{c}{\textbf{\textsc{Deletion}}} &
    \multicolumn{2}{c}{\textbf{\textsc{Insertion}}} &
    \multicolumn{2}{c}{\textbf{\textsc{Substitution}}}  \\
     ~ & ~ & EN & ZH & EN & ZH & EN & ZH \\
    \midrule
    \multirow{3}{*}{Google} & Devil’s Whisper~\cite{chen2020devil} & 33 & 21 & 4 & 0 & 3 & 1 \\
    
     ~ & NI-OCCAM~\cite{zheng2021black} & 25 & 27 & 8&2 &  6&3 \\ 
 	 
 	 ~ & TransAudio* & \textbf{80} & \textbf{29} & \textbf{11} & \textbf{3} & \textbf{6} & \textbf{5} \\
     
 	 \midrule
 	 
     \multirow{3}{*}{Tencent} & Devil’s Whisper~\cite{chen2020devil} & 46 & 22 & 7 & 1 & 0 & 0 \\
     ~ & NI-OCCAM~\cite{zheng2021black} & 53 &26& 19 & 6& 3& 5 \\
 	 
 	 ~ & TransAudio* & \textbf{78} & \textbf{27} & \textbf{26} & \textbf{18} & \textbf{3}& \textbf{6} \\
 	 \midrule
     \multirow{3}{*}{Alibaba} & Devil’s Whisper~\cite{chen2020devil} & 51 & 9 & 2 & 7 & 0 & 5 \\
     
     ~ & NI-OCCAM~\cite{zheng2021black} & 47 & 11& 6 &15& 4 & 16\\
 	 
 	 ~ & TransAudio* & \textbf{77}& \textbf{13} & \textbf{12} & \textbf{18} & \textbf{5} & \textbf{18} \\ 
    \bottomrule
    \end{tabular}}
\vspace{-0.2cm}
\caption{The attack success rate (SRoA) results of contextualized transfer-based attacks on commercial APIs. 'ZH' and 'EN' mean Chinese and English language respectively.}
\label{tab:spider}
\vspace{-0.6cm}
\end{table}

\paragraph{Evaluation on Cloud Speech APIs.} 
As presented in Table \ref{tab:spider}, we also compare the performance results of TransAudio$^*$ on three commercial ASR platforms.
We observe that TransAudio* is more capable of handling real-world attack scenarios, our method achieves higher SRoAs compared with other methods, and we attribute this to the two-stage learning of centralized AEs.
Interestingly, we find Google ASR system shows more vulnerability on deletion and substitution tasks, and the SRoA of insertion attack is highest on the Tencent ASR system.  

\section{Conclusion and Future Work}
In this paper, we propose an effective transfer-based targeted attack, namely TransAudio, against Automatic Speech Recognition (ASR) systems. With three contextualized perturbation behaviors, Deletion, Insertion, and Substitution, TransAudio can craft successful audio adversarial examples through a well-designed two-stage attack strategy.
Moreover, we propose an Audio Score-Matching optimization item to enhance the surrogate model for boosting the transfer-attack efficiency. This investigation of ASR model vulnerability could enhance interpretability and defense development, and future work will focus on building more robust speech recognition systems.

\vfill\pagebreak

\bibliographystyle{IEEEbib}
\bibliography{main}

\end{document}